\documentclass[aps,prd,twocolumn,10pt]{revtex4-1}

\usepackage{array}
\usepackage{graphicx}
\usepackage{amsmath}
\usepackage{hyperref}
\usepackage{mathrsfs}
\usepackage{breakurl} 

\newcommand\e[1]{\times10^{#1}}
\newcommand\td[2]{\frac{d#1}{d#2}}
\newcommand\then\Rightarrow
\newcommand\veps\varepsilon

\newcommand{\beq}[1]{\begin{equation}\label{#1}}
\newcommand{\eeq}{\end{equation}}
\newcommand{\bea}{\begin{eqnarray}}
\newcommand{\eea}{\end{eqnarray}}
\newcommand{\ba}{\begin{array}}
\newcommand{\ea}{\end{array}}

\newcommand{\rf}[1]{(\ref{#1})}
\def\be{\begin{equation}}
\def\ee{\end{equation}}
\newcommand{\eps}{\epsilon}

\newcommand{\Ethr}{E_{\text{thr}}}
\newcommand{\sthr}{s_{\text{thr}}}
\newcommand{\sobs}{s_{\text{obs}}}
\newcommand{\sigBSM}{\sigma_{\rm BSM}}
\newcommand{\BSMsig}{\sigma^{\rm BSM}}
\newcommand{\sigSM}{\sigma_{\rm SM}}
\newcommand{\SMsig}{\sigma^{\rm SM}}
\newcommand{\rhoSM}{\rho_{\rm SM}}
\newcommand{\rarr}{\rightarrow}

\begin{document}

\title{Using Integral Dispersion Relations to Extend the LHC Reach for New Physics}
\author{Peter B. Denton}
\email[Email: ]{peterbd1@gmail.com}
\author{Thomas J. Weiler}
\email[Email: ]{tom.weiler@vanderbilt.edu}
\affiliation{Department of Physics and Astronomy, Vanderbilt University, Nashville, TN 37235, USA}
\date\today

\begin{abstract}
Many models of electroweak symmetry breaking 
predict new particles with masses at or just beyond LHC energies. 
Even if these particles are too massive to be produced on-shell at the LHC, 
it may be possible to see evidence of their existence through the use of integral dispersion relations (IDRs). 
Making use of Cauchy's integral formula and the analyticity of the scattering amplitude, 
IDRs are sensitive in principle to changes in the cross section at arbitrarily large energies. 
We investigate some models of new physics.  
We find that a sudden, order one increase in the cross section above new particle mass thresholds 
can be inferred well below the threshold energy. 
On the other hand, for two more physical models of particle production, we show that 
the reach in energy and the signal strength of the IDR technique is greatly reduced.
The peak sensitivity for the IDR technique is shown to occur when the new particle masses are near the machine energy, 
an energy where direct production of new particles is kinematically disallowed, phase-space suppressed, or if applicable, 
suppressed by the soft parton distribution functions.
Thus, IDRs do extend the reach of the LHC, but only to a window around $M_\chi\sim \sqrt{s_{_{\rm LHC}}}$.
\end{abstract}

\maketitle

\section{Introduction}
Despite some predictions of a quick jump to new physics at the LHC~\cite{Abdullin:1998pm}, 
it seems distinctly possible that the next energy
scale for new physics is out of the reach of direct observation (or does not manifest itself as missing energy) at the LHC. 
It is, however, still possible to constrain certain new physics models at energies beyond those accessible at the LHC, 
by using integral dispersion relations (IDRs)~\cite{Block:1984ru}.

We briefly overview IDRs in section~\ref{sec:theory}. 
In section~\ref{sec:expoutlook} we discuss the present state and future expectations of LHC experiments 
in relation to the IDR technique.
We first solve these IDRs analytically in section~\ref{sec:SimpleModel} to understand their general behavior in certain limits.
We state the present status of total cross section parameterizations in section~\ref{sec:SM}, 
based on Standard Model (SM) assumptions.
In section~\ref{sec:newphysics}, we model how new physics beyond the SM and beyond the 
direct reach of accelerators may increase the cross section.
We consider one simple and two more physical model enhancements of the cross section.
In section~\ref{sec:results}, we discuss the reach of the IDR technique,
as illustrated with the various new physics models. 
Section~\ref{sec:results} ends with our conclusions and a brief outlook to the future.
Some details are presented in two appendices.

\section{Integral Dispersion Relation Theory}
\label{sec:theory}
A brief introduction to IDRs follows. For a more thorough introduction see reference~\cite{Block:1984ru}. 
The mathematics behind IDRs is
Cauchy's integral formula
\begin{equation}
f(z')=\frac1{2\pi i}\oint_{\partial A}\frac{f(z)}{z-z'}dz\,,
\label{eq:cauchy}
\end{equation}
for an analytic function $f$.
Here, the integration contour is around the boundary $\partial A$, $f(z)$ is analytic in the region $A$, 
and $z'\in A$. Next, we have the optical theorem
\begin{equation}
\sigma_{\text{tot}}=\frac{4\pi}p\Im f(\theta=0)\,,
\end{equation}
\begin{figure}[tb]
\includegraphics[height=2.2in]{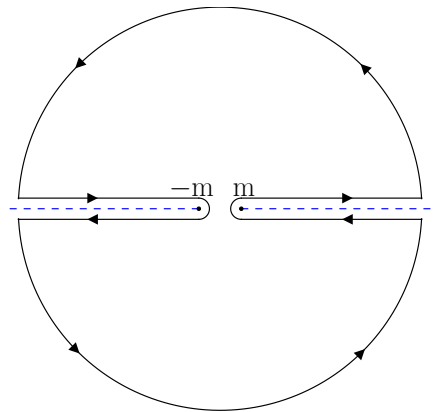}
\caption{The contour in the complex $E$ plane for the integral of $\mathscr F$ is shown.
Note that the physical $pp$ amplitude
approaches the right-hand cut from above and the $p\bar p$ amplitude approaches the left-hand cut from below.}
\label{fig:complex}
\end{figure}
which relates the  total cross section $\sigma_{\text{tot}}$ 
to the imaginary part $\Im f$ of the forward elastic scattering amplitude;
$p$ and $\theta$ are the center of mass (CoM) momentum and scattering angle respectively. 
Note that $\theta=0$ for elastic scattering is the same as $t=0$, 
where $t$ is the usual Mandelstam variable for the square of the transfer energy. 
The ratio of the real to the imaginary parts of the forward scattering amplitude is conventionally given the symbol
\begin{equation}
\rho(E)=\frac{\Re f(E,t=0)}{\Im f(E,t=0)} \,,
\end{equation}
where $E$ is the laboratory energy for a fixed target experiment,
related to the CoM energy squared ($s$) by $s=2m_p E+2m_p^2$.
Next, we select a particular closed curve shown in Fig.~\rf{fig:complex} in the complex plane with $R\to\infty$. We
then integrate $\mathscr F$, a complex valued function that is the analytic extension of the scattering amplitudes given by
\begin{equation}
f_{pp,p\bar p}(s,t=0)=\lim_{\veps\to0}\mathscr F(\pm(s+i\veps),t=0)
\end{equation}
around this contour.
In order to deal with convergence as the linear portions go to infinity for the actual behavior of the $pp,p\bar p$ cross sections, one
must also perform a subtraction leading to an additional
$f(E=0)$ constant. This gives the following equations for the forward scattering amplitudes
$f_{pp},f_{p\bar p}$ ($t=0$ now suppressed from our notation):
\begin{subequations}
\begin{multline}
\Re f_{pp}(E)=\Re f_{pp}(0)\\
+\frac E{4\pi^2}\mathcal P\int_m^\infty dE'\frac{p'}{E'}\left[\frac{\sigma_{pp}(E')}{E'-E}-\frac{\sigma_{p\bar p}(E')}{E'+E}\right]
\label{eq:singlesubtractionspp}
\end{multline}
\begin{multline}
\Re f_{p\bar p}(E)=\Re f_{p\bar p}(0)\\
+\frac E{4\pi^2}\mathcal P\int_m^\infty dE'\frac{p'}{E'}\left[\frac{\sigma_{p\bar p}(E')}{E'-E}-\frac{\sigma_{pp}(E')}{E'+E}\right]
\label{eq:singlesubtractionsppbar}
\end{multline}
\label{eq:singlesubtractions}
\end{subequations}
where $\sigma$ is the total cross section, $(\mathcal P\int)$ is the
principal value integral, and
$p'\equiv\sqrt{E'^{\,2}-m_p^2}$.
By the Pomeranchuk theorem~\cite{Pomeranchuk}
(and supported by QCD-parton considerations and experimental evidence)
$\Delta\sigma=\sigma_{pp}-\sigma_{p\bar p}\to0$ as $E\to\infty$, thus guaranteeing convergence 
of the integrals in Eqs.~(\ref{eq:singlesubtractionspp},\ref{eq:singlesubtractionsppbar});
were these integrals still divergent, doubly subtracted IDRs would be necessary. 

Finally, these formulas together with the optical theorem again give the useful IDRs
\begin{subequations}
\begin{multline}
\rho_{pp}(E)\sigma_{pp}(E)=\frac{4\pi}p\Re f_{pp}(0)\\
+\frac E{p\pi}\mathcal P\int_m^\infty dE'\frac{p'}{E'}\left[\frac{\sigma_{pp}(E')}{E'-E}-\frac{\sigma_{p\bar p}(E')}{E'+E}\right]\,,
\label{IDR}
\end{multline}
\begin{multline}
\rho_{p\bar p}(E)\sigma_{p\bar p}(E)=\frac{4\pi}p\Re f_{p\bar p}(0)\\
+\frac E{p\pi}\mathcal P\int_m^\infty dE'\frac{p'}{E'}\left[\frac{\sigma_{p\bar p}(E')}{E'-E}-\frac{\sigma_{pp}(E')}{E'+E}\right]\,.
\label{IDRbar}
\end{multline}
\end{subequations}

A built-in assumption from Eq.~\rf{eq:cauchy} is that the scattering amplitude is analytic. Based on the fact that current
fits to the inelastic and total cross sections suggest that the proton asymptotically approaches a black disk, the evidence
strongly suggests that the $pp$ scattering amplitude is analytic~\cite{Block:2012nj}.

In addition, the forward scattering amplitudes $f_{pp}(0)=f_{p\bar{p}}(0)$ are consistent with being purely imaginary at the
non-physical energy $E=0$ so the subtraction constant is ignorable~\cite{Block:2005ka}.

In what follows, we exploit Eq.~\rf{IDR} to extend the reach of the LHC to energies beyond 
the LHC-equivalent fixed-target energy $E\equiv (s-2m_p^2)/2m_p$.
Formally, the IDRs relate contributions from new physics occurring all the way up in energy to infinity,
to observables at present energies.
In practice, the integral of the IDR falls off in energy away from the observation energy,
so the reach beyond present energies will be limited.

\quad

The general strategy that we will use to explore new physics is to first use the IDR to calculate $\rho$ at a particular energy 
(an LHC energy) without the inclusion of new physics.
Then we calculate $\rho$ at the same energy with the inclusion of the new physics cross section. 
Since $\rho$ can be calculated without IDRs in a model-independent fashion, as briefly described in
appendix~\ref{appx:rhocalculation}, enhancements of the cross section can be either identified or ruled out
by comparing theoretical and experimental values of $\rho(E)$.

\section{Experimental Status}
\label{sec:expoutlook}
The TOTal Elastic and diffractive cross section Measurement (TOTEM~\cite{TOTEM}) experiment at the LHC 
is designed to measure forward cross sections by probing very low $|t|$ regions. 
TOTEM places a series of Roman pot detectors very close to the beam and very far from the interaction point. 
With improved LHC optics, 
TOTEM should be able to provide an improved measurement of $\rho$ independent  of IDRs~\cite{Antchev:2011vs}. 
A comparison of TOTEM's $\rho$, so determined, with the IDR prediction of $\rho$,
then provides the potential evidence for new physics.

Similarly, the Absolute Luminosity For the ATLAS (ALFA~\cite{ATLAS-CONF-2010-060}) experiment, the
LHC forward (LHCf~\cite{LHCf}) experiment, along with a host of others will also make comparable measurements 
in an attempt to improve the precision of the luminosity calculation, which is necessary to infer $\sigma_{\text{tot}}$,
and then to infer $\rho$ without the use of IDRs~\cite{Aad:2011dr} (see appendix \ref{appx:rhocalculation} below).
Thus, there are several experiments that aim to measure the total cross section. These offer hope for smaller error bars on
IDR-independent determinations of the crucial parameter $\rho(E)$.

A recent $\sqrt s=7$ TeV TOTEM paper presented a state of the art value for the IDR-independent $\rho$, of $\rho=0.145$
with error bars of $\sim60\%$~\cite{TOTEM7}. TOTEM cited a 95\% significance level 
(roughly speaking, a $2\sigma$ bound) that $\rho<0.32$. 
Comparing this to the SM prediction of $\rho(\sqrt s=7$ TeV$)=0.1345$
gives an upper limit of the fractional increase $(\rho-\rhoSM)/\rhoSM=1.38$ at the 95\% significance level.
For brevity, in what follows we denote $(\rho-\rhoSM)/\rhoSM$ as $\Delta\rho/\rho$.

As an illustrative example of what a future determination of $\rho$ might mean for the IDR technique,
we investigate a definite value for $\rho$; 
we choose as the definite value the experimentally-inferred mean value $\rho(\sqrt s=7$ TeV$)=0.145$.
For this example, the fractional increase in $\rho$ is $\Delta\rho/\rho=0.0781$.
This value for $\rho$ is chosen for illustration only, as it offers insight into the merit of IDRs should 
experiments greatly reduce their errors in the inference of $\rho$.
The chosen value nicely exceeds the SM prediction by 7\%, 
but with almost zero significance $\sim0.1\sigma$ at present.

\quad

To get a feel for the reach and nature of this integral dispersion relation approach, 
we examine the integral under simplifying approximations in the next section.
Then in the following sections, we examine the SM contribution to the IDRs and $\rho$,
and the contributions from three constructed models of new physics.

\section{A Simplified Dispersion Integral to Set Expectations}
\label{sec:SimpleModel}
In this section we make two assumptions to reduce the dispersion integrals in Eqs.~\rf{IDR} and~\rf{IDRbar} to a form 
that can be integrated analytically.
While neither assumption is strictly valid, they are useful to reveal the gross features of the dispersion integral.
The first assumption is to set $m_p$ to zero.  Besides replacing the lower limit of integration with zero, this
assumption also sets $p'/E'$ equal to one.  The second assumption is 
to set $\sigma_{pp}$ and $\sigma_{p\bar{p}}$ equal to each other, and to a constant which we call $\sigma_0$.
With these two assumptions, both dispersion integrals can be written as 
\beq{simple1}
(2\sigma_0)\,{\cal P} \int \frac{dx}{x^2-1}=(2\sigma_0)\,\log\left(\frac{|1-x|}{1+x}\right)\,,
\eeq
with $x\equiv E'/E$ and is
valid everywhere except at $x=1$, where the integral is singular.
Blind evaluation of the definite integral over the range $[0,\infty]$ then gives zero.
That this is correct can also be seen in the following way:
By definition, the definite integral from Eq.~\rf{simple1} is 
\beq{simple2}
\lim_{\eps\rarr 0} \left[ \int_0^{1-\eps} \frac{dx}{x^2-1} + \int_{1+\eps}^\infty \frac{dx}{x^2-1}\right]\,.
\eeq
Replacing $x$ by $u\equiv\frac{1}{x}$ in either integral, maps the integration region into that of the 
other integral, and \emph{reveals that the two integrals are equal but with opposite sign}. 
Thus, the total integral vanishes.
In particular, the singularity in the integrand vanishes in the principal value.

In Fig.~\rf{fig:Integrand_xmin}, we plot the integrand $(x^2-1)^{-1}$ of our simplified dispersion integral. 
As the lower limit of integration $x_{\min}$ is moved up from zero, 
the cancellation above and below the singularity is no longer complete.
However, the vanishing of the total integral when integrated from from zero to infinity
allows us replace the integration across the singularity with a simple, manifestly nonsingular integral as follows:
\beq{simple3}
{\cal I}(x_{\min})\equiv \int_{x_{\min}}^\infty \frac{dx}{x^2-1}= \int_0^{x_{\min}} \frac{dx}{1-x^2} \,.
\eeq
For $x_{\min}=1$, the cancellation is maximally incomplete and the integral is infinite.
We plot ${\cal I}(x_{\min})$ in Fig.~\rf{fig:Integral_xmin}.
As expected, the integral is everywhere positive, and diverges at $x_{\min}=1$. The divergence seems unphysical in that it corresponds
to either $\Im f=0\then\sigma_{tot}=0$ by the optical theorem which shouldn't be the case or that $\Re
f\to\infty\then\sigma_{tot}\to\infty$ which is also unphysical. Since these particles have mass (which is ignored here) they have a
finite lifetime and a finite width which would keep this integral finite at $x_{\min}=1$.

\begin{figure}
\centering
\includegraphics[width=\columnwidth]{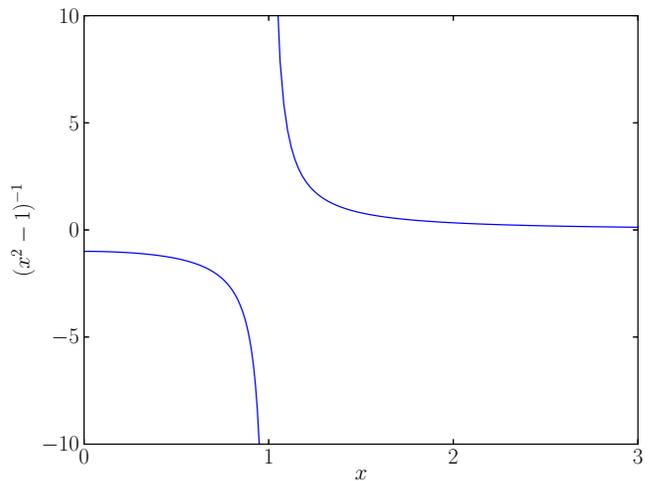}
\caption{The integrand of the IDRs with the $\sigma=$ constant and $m_p\to0$ limits taken.}
\label{fig:Integrand_xmin}
\end{figure}

\begin{figure}
\centering
\includegraphics[width=\columnwidth]{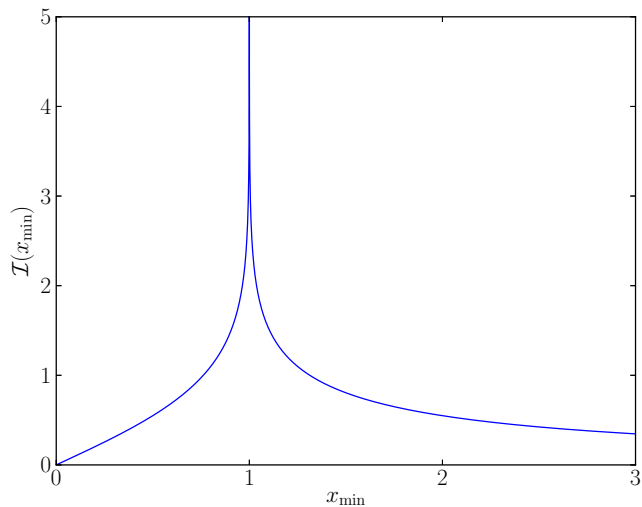}
\caption{The integral of the IDRs with the $\sigma=$ constant and $m_p\to0$ limits taken. 
We see the expected singularity at $x_{\min}=1$.
For the new physics contribution, $x_{\min}=\Ethr/E$.}
\label{fig:Integral_xmin}
\end{figure}

We may ask how the singularity is approached, from below and from above.
Writing $x_{\min} =1-\Delta$ and~$1+\Delta$, we have the two integrals
$\int_{1-\Delta}^\infty \frac{dx}{x^2-1}$ and $\int_{1+\Delta}^\infty \frac{dx}{x^2-1}$.  
The first integral crosses the singularity and 
according to Eq.~\rf{simple3} is equal to the clearly finite integral $\int_0^{1-\Delta} \frac{dx}{1-x^2}$.
With the replacement $x\rarr 1/x$, the second integral becomes $\int_0^{\frac{1}{1+\Delta}} \frac{dx}{1-x^2}$.
Thus, the two integrations differ only in the upper limit of integration.  At first order in $\Delta$ they are identical,
as they must be to give a finite principal value integral.
At higher order in $\Delta$, the second integral exceeds the first integral.
So we expect ${\cal I}(x_{\min})$ to show symmetry about the singular value $x_{\min}=1$ for small deviations,
but a larger value above $x_{\min}=1$ than below for larger deviations.
This expectation is visible in Fig.~\rf{fig:Integral_xmin}.

Why do we investigate $x_{\min}$ values other than zero?
When new physics enters at a threshold energy $E_{\rm thr}$, the contribution of the new physics 
to the dispersion integral begins at $x_{\min}=\frac{E_{\rm thr}}{E}$, where $E$ is the energy of the accelerator.
Thus, Fig.~\rf{fig:Integral_xmin} gives the shape of the new physics contribution as a function 
of the new physics threshold. In what follows, our much more realistic parameterizations of new physics 
will present curves that qualitative have the form given by the simplistic model discussed in this section.

We may summarize this section by saying that the SM cross section is expected to give a modest contribution to the 
dispersion integral (zero in our simplistic model of constant and equal $pp$ and $p\bar{p}$
cross sections with vanishing proton mass).  On the other hand, new physics enters at a nonzero threshold
which implies an incomplete cancellation in the dispersion integral, and thus a possibly significant 
contribution to the dispersion integral.  Therefore, the ratio of new physics to total physics as revealed 
in the IDR potentially offers an observable window to new physics even with threshold energy above the 
direct reach of the LHC.

\section{\texorpdfstring{\boldmath $\rho(E)$}{rho(E)} in the Standard Model }
\label{sec:SM}
Before venturing into speculative constructions of new physics contributions,
we present the SM contribution.

In the real world, the exact cancellation of the integral presented in the previous section 
does not happen because our assumptions are (slightly) violated.  
In the real world, the proton mass $m_p$ is not zero,
and the $pp$ and $p\bar{p}$ cross sections are neither equal nor constant in energy.
Fits to data suggest that the cross sections 
decrease with $E$ until $E\sim 60$~GeV ($\sqrt s=10.6$~GeV) and $E\sim250$~GeV ($\sqrt s=21.8$~GeV) 
for $pp, p\bar p$ respectively, before increasing.
Furthermore, Froissart theory tells us that a $\log^2s$ growth eventually dominates the energy dependence, a fact that has been
confirmed with fits to present experimental data.

The SM total $pp,p\bar p$ cross section $\sigma_{SM}$ is typically parameterized as
\bea 
\sigma_{SM}(E) = c_0 &+& c_1\log\left(\frac Em\right)+c_2\log^2\left(\frac Em\right) \nonumber \\
    &+& c_3\left(\frac Em\right)^{-\frac12}\ \pm c_4\left(\frac Em\right)^{\alpha-1}
\eea
where $m_p$, the proton mass, is used as the energy scale. $c_i,\alpha$ are fit parameters with $\alpha<1$. 
The $E^{-\frac12}$ term is a result of invoking Regge behavior. 
The upper sign refers to $pp$ scattering and the lower to $p\bar p$ scattering. 
This form is motivated by being the most general and fastest rising form allowed by the Froissart bound. 
The values and precision of the $c_i$ and $\alpha$ from~\cite{Block:2005ka} are shown in table~\rf{tab:fit}.
The total $pp$ cross section for the SM is included in Fig.~\rf{fig:mods} (solid line, labeled as the $h_0$ case).

Note that different fits to the $pp,p\bar p$ cross section do not substantively change the results of this paper.
The current limits on the $pp$ total cross section are predominately derived from data at and below the LHC. Fits to
functions that behave differently than $\log^2(s)$ such as $\log(s)$ and $s^\epsilon$ have been essentially ruled out
\cite{Block:2004ek,Block:2012nj}. Auger does quote a value for the $pp$ total cross section at $57$~TeV
\cite{Collaboration:2012wt}, but the precision is low (a fractional error of $\sim0.35$) and depends on specifics built into the
Glauber model. It does not severely limit the high energy behavior of the cross section.

\begin{table}
\centering
\begin{tabular}{|l|l|}
\hline
$c_0$ (mb)	&$36.95$\\\hline
$c_1$ (mb)	&$-1.350\pm0.152$\\\hline
$c_2$ (mb)	&$0.2782\pm0.105$\\\hline
$c_3$ (mb)	&$37.17$\\\hline
$c_4$ (mb)	&$-24.42\pm0.96$\\\hline
$\alpha$	&$0.453\pm0.0097$\\\hline
\end{tabular}
\caption{Fit parameters~\cite{Block:2005ka} with various analyticity constraints.}
\label{tab:fit}
\end{table}

Concerning the first approximation of the previous section, namely $m_p=0$, 
we find that returning the physical, nonzero $m_p$ to the integral (including $p'/E'\ne 1$) 
gives nonzero but negligible integral values of $2.649\e{-8}$ and $5.966\e{-9}$ at LHC energies $\sqrt{s}=7$ and~14~TeV\@. 
On the other hand, keeping $m_p$ zero but returning to $\sigma_{pp}$ and $\sigma_{p\bar{p}}$ their realistic energy dependences 
yields nonzero integral values of $0.1345$ and $0.1309$ at $\sqrt{s}=7$ and~14~TeV\@.
And finally, using nonzero $m_p$  and realistic $pp$ and $p\bar{p}$ cross sections returns the values $0.1345$ and $0.1309$ at
$\sqrt{s}=7$ and~14~TeV\@.
The final two integration sets (realistic $\sigma$'s and zero or nonzero $m_p$) 
agree to about seven to eight decimal places respectively (on the order of $m_p^2/s$).
The conclusion is that the $m_p\to0$ approximation is generally a valid one,
whereas the constant and equal SM cross section approximation in the previous section is not.
However, the integral contributions of the SM to the IDRs (the solid lines in Fig.~\rf{fig:rhoVssobstr20}) are not large,
and we are encouraged to pursue further the contributions that might arise due to physics beyond the SM.

For the subtraction constant, we will take $f(0)=0$, since the value from the fits above is $f(0)=-0.073\pm0.67$~mb~GeV\@. 
We note that even at the value $1\sigma$~away from zero, the term $4\pi\,f(0)/(p\,\sigma_{pp} )$ at LHC energies 
contributes less than one part in $10^5$ to $\rho$.

\section{New Physics Contributions}
\label{sec:newphysics}
We turn now to the construction of three models for new physics beyond the Standard Model (BSM).
We consider a class of modified cross sections of the general form
\begin{equation}
\sigma(s)=\sigma_{SM}(s)[1+h_i(s)]
\label{eq:sigma_mod}
\end{equation}
where the $h_i=(\sigBSM/\sigSM)_i$ are cross section ratios;
they vanish below the threshold $\sthr$ for new physics.
We apply the same enhancement to both $\sigma_{pp}$ and $\sigma_{p\bar p}$ since, by the Pomeranchuk
theorem discussed above, each cross section should respond to new physics in the same way at energies well above the
proton mass. 

The first model we present is described by a simple step function at $\sthr$.
This model results in an especially  close analogy to the idealized IDRs we discussed in section \ref{sec:SimpleModel}.
In particular, due to its nonzero new cross section at $E=\Ethr$, 
this model yields a singularity in the IDR integrand at $E=\Ethr$ and therefore a singular value for $\rho(E=\Ethr)$.

More realistically, we expect phase space to present a cross section for new physics that has no jump discontinuity at threshold.  
For example, two-body phase space is $\beta/8\pi$, where $\beta$ is either particle velocity in the CoM frame;
at threshold, $\beta$ is identically zero.
Furthermore, including parton distribution functions to the model also yields a zero cross section at threshold.
The new physics matrix elements may also vanish right at threshold.
So we are led to the next two models of BSM physics.
The second model we present involves hard-scattering parton production of new particles,
while the third model is constructed from diffractive phenomenology.
The second and third models provide cross sections that vanish at threshold, 
leading to finite values for $\rho(E=\Ethr)$.

Since only the first model, the step function, yields a nonzero change in the cross section at threshold, 
the $\rho$-value resulting from model $h_1$ should be considered an upper bound to the contribution of 
new physics BSM.  The bounding of cross sections by the $h_1$ step function model is evident 
in Fig.~\rf{fig:mods}, where we show the SM cross section (given by zero enhancement and labeled by $h_0=0$) 
and its enhancements ($h_i,\ i=1,2,3$) by the three new physics models that are presented in detail below.

\begin{figure}[tb]
\centering
\includegraphics[width=\columnwidth]{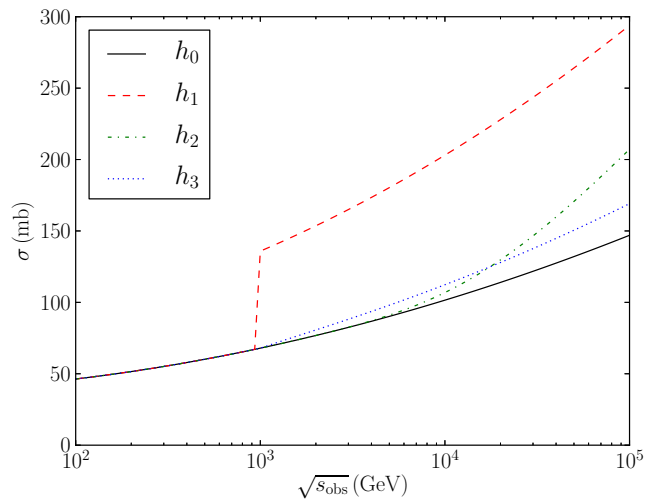}
\caption{The $pp$ total cross sections $\sigSM$~+~$\sigBSM$ are plotted, 
for the choice $\sqrt{\sthr}=M_\chi=10^3$ GeV\@. 
Doubling of the cross section at threshold is assumed for the $h_1$ model, i.e., $D=1$.
The SM cross section uses the parameters~\cite{Block:2005ka} shown in Table~\rf{tab:fit}. 
The slow initial rise in $h_2$ is a result of the parton distribution functions;
the asymptotic value of $h_2$ is determined by $\sigma_{\text{inel}}/\sigma_{\text{tot}}$. 
While $\lim_{s\to\infty}h_3$ is small compared to other models, it rises quickly at threshold, in contrast to $h_2$.}
\label{fig:mods}
\end{figure}

No new conserved quantum number is assumed in our models 
(valid, e.g., for broken $R$-parity SUSY models).
Thus, energy is the only impediment to production of heavy new single particles,
and the heavy single mass value $M_\chi$ determines $\Ethr$.
Without a new quantum number, the new particle would decay to SM particles, 
and due to its large mass, decay very quickly. 
Consequently, other than invariant mass combinatorics, there is no good signature of the new particle's production.
One may have to rely on IDRs and/or an anomalous $\Delta\sigma$ for new particle identification.
Thus, we plot $\Delta\rho/\rho$ and
$\Delta\sigma/\sigma$ versus $M_\chi$ ($M_\chi=\sqrt{\sthr}-2m_p\approx\sqrt{\sthr}$), 
to see if the IDR technique can identify new physics via an anomalous $\rho$ measurement,
before the new physics would be directly noticeable in the cross section increase.

\subsection{A Simple BSM Enhancement -- a Step Function}
\label{subsec:asimpleenhancement}
A simple example function for BSM physics is 
\begin{align}
h_1(s)&=D\,\Theta(s-\sthr)\,,
\label{eq:mods}
\end{align}
a step-wise jump in cross section at the threshold CoM energy $\sqrt{\sthr}$. 
The parameter $D$ is a measure of the size of the new cross section relative to the SM. 
For example, in the event that new physics exactly doubles the cross section at
$s=\sthr$, then we take $D=1$.
Modifying $\sigma_{SM}$ in the form of Eq.~\rf{eq:mods} guarantees that the new total cross section
$\sigma$ continues to grow as fast as $\sigma_{SM}\propto\log^2s$ (but not faster), 
and that the new physics contribution remains large over a sizable energy range
beyond the threshold energy.

As mentioned above, an unphysical aspect of the step function enhancement is a non-vanishing cross section at threshold, 
which leads to the uncanceled singularity in the IDR integrand at $E=\Ethr$ and a singular value for $\rho(E=\Ethr)$.
However, the model has redeemable value in that the width of the singularity is small.
Thus, the model offers a meaningful upper bound to new particle production away from the singularity.

\subsection{A Partonic Model of New Particle Production}
\label{subsec:amorephysicalmodel}
The most popular model of new physics at the electroweak symmetry breaking scale is $R$-parity preserving supersymmetry (SUSY),
with masses tuned to the EW-scale to stabilize the ratio $m_h/M_{\rm Planck}$ (the ``hierarchy problem'').
Unfortunately, $R$-parity conservation requires $s$, $t$, and $u$ to have EW-scale values, which severely
suppresses the SUSY cross section to about $10^{-10}$ times the SM cross section.
However, as LHC limits on $R$-parity conserving SUSY are becoming more constraining, 
$R$-parity violating (RPV) models are getting a closer look
~\cite{Bazzocchi:2012ve,Bhattacherjee:2013gr,Wei:1900zz,Chatrchyan:2013xsw,Cakir:2013toa}.
If $R$-parity is violated, we can replace one final state particle from a SM process with an effectively identical heavier
counterpart for each possible final state. 
Then the only difference between the modified cross section and the SM cross section comes in the
form of the reduced final state phase space and the threshold parton energy.
Importantly, the fast growing $\log^2s$ contribution to the SM $\sigma_{\text{tot}}$, which arises from soft and collinear gluon
divergences, may be maintained.
Also, other exotic models with extra dimensions~\cite{Chialva:2005gt}
and a non-conserved KK number might grow a large cross section as a power law instead of the Froissart $\log^2s$ limit.

Let $\sigma_i(s)=\sigma_i(pp\to\dots)$ be the SM cross section and $\BSMsig_i(s)=\sigma_i(pp\to\chi+\dots)$ be the new physics
contribution, where $i=\{$el, inel, tot$\}$, and dots denote additional SM particles in the final states. Hats will denote parton cross sections
instead of $pp$ cross sections. We note that since $\BSMsig_{\text{el}}=0$, then $\BSMsig_{\text{inel}}$ must equal
$\BSMsig_{\text{tot}}$. 
Then the physical total $pp$
cross section is $\sigma_{\text{tot}}+D\,\BSMsig_{\text{inel}}=\sigma_{\text{tot}}(1+D\,h_2(s) )$ where
$h_2=\sigma_{\text{inel}}^{\rm BSM}/\sigma_{\text{tot}}$ in the form of Eq.~\rf{eq:sigma_mod}.

We start with an expression of the conservation of momentum for the new physics contribution.
\begin{equation}
\BSMsig_{\text{tot}}(s)=\sum_{i,j}\int_{(\hat s>M_\chi^2)} dx_1dx_2f_i(x_1)f_j(x_2)\hat\sigma^{\rm BSM}_{\text{tot}}(\hat s)
\label{eq:p conservation}
\end{equation}
where $\hat s\approx x_1x_2s$ is the parton CoM energy and the $f_i$ are the various parton distribution functions~(pdfs). 
Let the SM final state masses be zero. 
The summations are over parton types and the integrals are over the accessible $x_1,x_2$ space: $\hat s>M_\chi^2$.

If we assume that for each SM particle in the final state, there is an analogous new particle $\chi$ produced with the same coupling, 
then there is little $t$- or $u$-channel propagator suppression (see appendix \ref{appx:mint}), and so 
the matrix elements will be similar.
The new, heavier final state masses suppress only the available phase space.
So we can set parameter $D=1$, and write
\begin{equation}
\frac{\hat\sigma^{\rm BSM}_{\text{tot}}(\hat s)}{\sqrt{\lambda(\hat s,M_\chi^2,0)}}=\frac{\hat\sigma_{\text{inel}}(\hat s)}{\sqrt{\lambda(\hat
s,0,0)}}
\label{eq:sm to dm}
\end{equation}
where the triangle function (symmetric in its arguments) is defined as 
\begin{equation}
\lambda(a,b,c)=a^2+b^2+c^2-2ab-2bc-2ca \,.
\end{equation}
The inelastic cross section shows up in the SM-equivalent case since the related new particle cross sections 
must be inelastic. 

It is easy to see that the relevant ratio can be simplified to
\begin{equation}
\sqrt{\frac{\lambda(\hat s,M_\chi^2,0)}{\lambda(\hat s,0,0)}}=1-\frac{M_\chi^2}{\hat s}
\label{eq:sqrt lambda}
\end{equation}
Then, combining Eqs.~(\ref{eq:p conservation},\ref{eq:sm to dm},\ref{eq:sqrt lambda}), 
and integrating out the internal $\sigma_{\text{tot}}(\hat s)$ which leaves behind a factor of $\hat s/s=x_1x_2$,
we obtain
\begin{multline}
\BSMsig_{\text{inel}}(s)=\sigma_{\text{inel}}(s)\sum_{i,j}\int dx_1dx_2\\
\times f_i(x_1)f_j(x_2)x_1x_2\left(1-\frac{M_\chi^2}{\hat s}\right)\,.
\end{multline}

Consistency of this model derivation gains support by noting that 
as either $M_\chi\to0$ or $s\to\infty$, we recover the SM cross section 
(recalling that $\sum_i\int dxf_i(x)x=1$ expresses conservation of momentum 
when the momentum of the parent nucleon is partitioned among partons).

Let us introduce the ratio $z\equiv\sigma_{\text{inel}}/\sigma_{\text{tot}}$. 
As suggested by the black disk limit,  $z\to\frac12$ as $s\to\infty$.
However, data for the LHC $\sqrt s=7$ TeV run, and CR data in the vicinity of $\sqrt s=57$ TeV
suggest that $z$ is well (and conservatively) approximated as a constant $z\approx0.7$~\cite{Block:2012nj}.
Our interest is the upcoming $\sqrt s=14$~TeV LHC run, for which $z\approx0.7$ is the appropriate value.
Finally, we arrive at our model for new physics:
\begin{multline}
h_2(s,M_\chi)=z\sum_{i,j}\int_{x_1x_2>M_\chi^2/s}dx_1dx_2\\
\times f_i(x_1,M_\chi)f_j(x_2,M_\chi)x_1x_2\left(1-\frac{M_\chi^2}{\hat s}\right)\,.
\label{eq:h2}
\end{multline}
Of course, the parton distribution functions $f_i$ also depend on the transfer energy $Q$, which we 
take to be $M_\chi$.  
For our numerical work with pdfs, we use the CT10 parton distribution functions~\cite{Gao:2013xoa}.

Note that this model has a vanishing cross section right at threshold, (at $\hat{s}=M^2_\chi$),
due to the $(1-M^2_\chi/\hat{s})$ factor, and due to the vanishing parton distributions at threshold.
Thus, $\rho$ is finite for all $E$ values, including the peak at $E=\Ethr$.
Furthermore, the rise from threshold is the very slow,
a notable feature of the $h_2$ model.
This slow rise in $h_2$ is evident in Fig.~\rf{fig:mods}.
We find that the slow rise is due to the suppressed pdfs near threshold;
the phase-space reduction factor contributes a negligible suppression to the rise.
We conclude that any deep inelastic model with partons as initial state particles 
will experience a similar slow rise from threshold.
Finally, we note from Eq.~\rf{eq:h2} that $h_2$ has a finite asymptotic value of $z\approx 0.7$,
i.e., about a 70\% increase over the SM cross section.

\subsection{A Diffractive Model of New Particle Production}
\label{subsec:nonparton}
An alternative to the partonic approach just presented is to consider general descriptions 
of $pp$ inelastic cross sections without reference to partonic substructure. 
Inelastic cross sections can be described by the parameter $\xi\equiv M_X^2/s$. 
$M_X$ is defined by first making a pseudorapidity ($\eta$) cut at the mean
$\eta$ of the two tracks with the greatest difference in $\eta$. $M_X$ is then taken as the larger invariant mass of the two halves. 
Ref.~\cite{Aad:2011eu} provides a model form for the inelastic cross section.
It is
\begin{equation}
\td\sigma\xi\propto\frac{1+\xi}{\xi^{1+\epsilon}}
\label{eq:dsigmadxi}
\end{equation}
where $\epsilon=\alpha(0)-1$ and $\alpha(0)$ is the Pomeron trajectory intercept at $t=0$. 
Values for $\epsilon$ are typically in the $[0.06,0.1]$ range. 
We take the mean of this range, $\epsilon=0.08$, in this paper. 
Next, we note that $1\ge\xi>m_p^2/s\equiv\xi_p$, since
$\xi_{\min}=\xi_p$ describes elastic scattering. 
To find the total cross section, we integrate Eq.~\rf{eq:dsigmadxi} across
$\xi\in[\xi_p,1]$ and get
\begin{equation}
\sigma\propto\frac{(1-2\epsilon)+(\epsilon-1)\xi_p^{-\epsilon}+\epsilon\xi_p^{1-\epsilon}}{\epsilon\,(\epsilon-1)}
\label{eq:sigmaximin}
\end{equation}

As an interesting aside, we note that to order $\epsilon^1$ in Eq.~\rf{eq:sigmaximin},
the leading energy behavior grows like $\log^2(s/m_p^2)$,
thereby providing the expected asymptotic Froissart growth~\cite{note1}. 
However, higher order terms in $\epsilon$ lead to higher logarithmic orders,
indicating that Eq.~\rf{eq:sigmaximin} is pre-asymptotic.

We now consider a rapidity cluster containing a new particle of mass $M_\chi$. 
With the substitution $\xi_p\to\xi_\chi\equiv M_\chi^2/s$ in  Eq.~\rf{eq:sigmaximin}, divided by the SM case,
 we arrive at the useful ratio 
\begin{equation}
R(M_\chi,s)\equiv\frac{\BSMsig_{\rm diff}}{\SMsig_{\rm diff}}=\frac{1-2\epsilon+(\epsilon-1)\xi_\chi^{-\epsilon}+\epsilon\xi_\chi^{1-\epsilon}}{1-2\epsilon+(\epsilon-1)\xi_p^{-\epsilon}
+\epsilon\xi_p^{1-\epsilon}}
\end{equation}
Next we note the relation in Eq.~\rf{eq:dsigmadxi} describes single dissociative processes, which constitute
only 15\% of the inelastic cross section. 
We make the model assumption that the remaining 85\% of the inelastic cross section, including
double dissociative and non-diffractive
processes, are also governed by the form in Eq.~\rf{eq:dsigmadxi}.
Finally, we include the factor $z=\sigma_{\text{inel}}/\sigma_{\text{tot}}\approx0.7$ described in the previous subsection,
and make explicit the on-shell requirement $M^2_\chi\le s$ with a Heaviside function, to arrive at our final 
model expression
\begin{equation}
h_3(s)=z\,\frac{1-2\epsilon+(\epsilon-1)\xi_\chi^{-\epsilon}+\epsilon\xi_\chi^{1-\epsilon}}{1-2\epsilon+(\epsilon-1)\xi_p^{-\epsilon}
+\epsilon\xi_p^{1-\epsilon}}\;\Theta(1-\xi_\chi) \,.
\end{equation}
As with model $h_2$, model $h_3$ has the desirable feature that the BSM cross section vanishes at 
threshold (here, $\xi_\chi =1$).  Thus, $\rho$ is finite all energies, including the peak at $E=\Ethr$.

In Fig.~\rf{fig:mods} we see that the $h_3$ model rises more quickly at threshold than the $h_2$ model, 
but attains a smaller asymptotic value:
\begin{equation}
\lim_{s\to\infty}h_3(s)=z\left(\frac{m_p}{M_\chi}\right)^{2\epsilon}\approx0.23\left(\frac{1\text{ TeV}}{M_\chi}\right)^{2\epsilon}\,,
\label{eq:h3limit}
\end{equation}
i.e., about a 25\% increase beyond the SM cross section.
This faster rise but lower asymptotic value for $h_3$ compared to $h_2$ is evident in Fig.~\rf{fig:mods}. 

\section{Results}
\label{sec:results}
For each of the three models discussed in the previous section, 
we calculate the effect they have on $\rho$. 
The parameter considered is the fractional increase in $\rho$, 
given as $\Delta\rho/\rho\equiv (\rho-\rhoSM)/\rhoSM$. 
This is then related to the TOTEM results at $\sqrt s=7$~TeV\@. 
We consider the mean value from their experiment as an example signal: 
$\rho=0.145$ ($\pm0.091$, $1\sigma$ confidence level) is compared to the SM prediction 
of $\rho=0.1345$, a value which implies a fractional increase of $\Delta\rho/\rho=0.0781$. We also look at the 
TOTEM upper limit, given as $\rho<0.32$ at the $2\sigma$ confidence level, 
leading to a maximum fractional increase of $\Delta\rho/\rho<1.38$ ($2\sigma$).

\subsection{Results From the Step Function Model}
The step function enhancement of $\rho$ is shown in Fig.~\rf{fig:rhoVssobstr20}.
As expected, a shape very similar to that  of Fig.~\rf{fig:Integral_xmin} results.
For comparison, the SM behavior of $\rho$ is also shown.
Here we have taken $D=1$ and $\sqrt{\sthr}=20$~TeV\@.
We see for a doubling of the cross section at $\sqrt{\sthr}=20$~TeV, 
a small increase in $\rho$ is evident already at an energy an order of magnitude below $\sqrt{\sthr}=20$~TeV,
and that $\rho$ increases by nearly a factor of four at $\sqrt{s_{\text{obs}}}=14$~TeV\@.

\begin{figure*}
\centering
\includegraphics[width=\columnwidth]{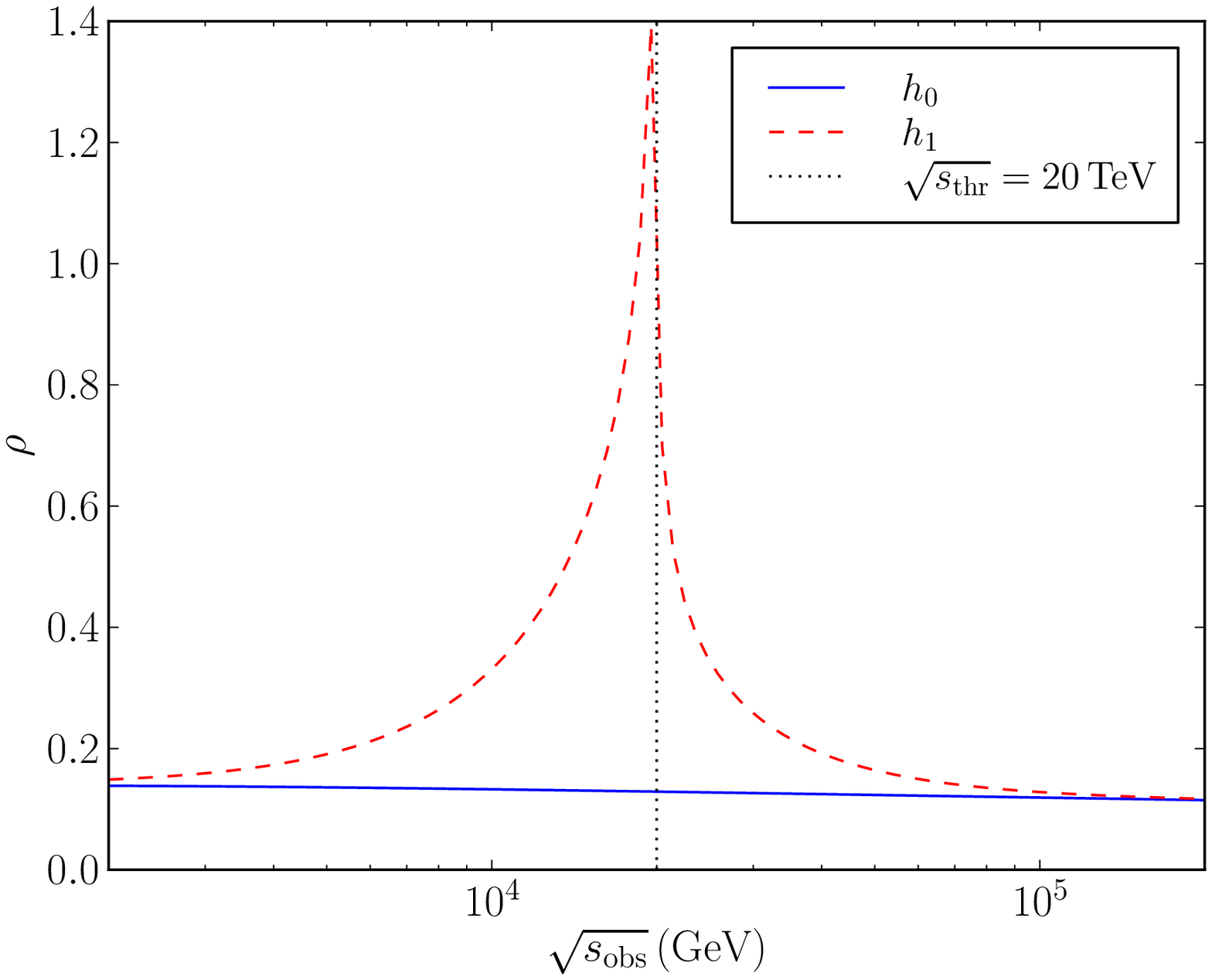}
\includegraphics[width=\columnwidth]{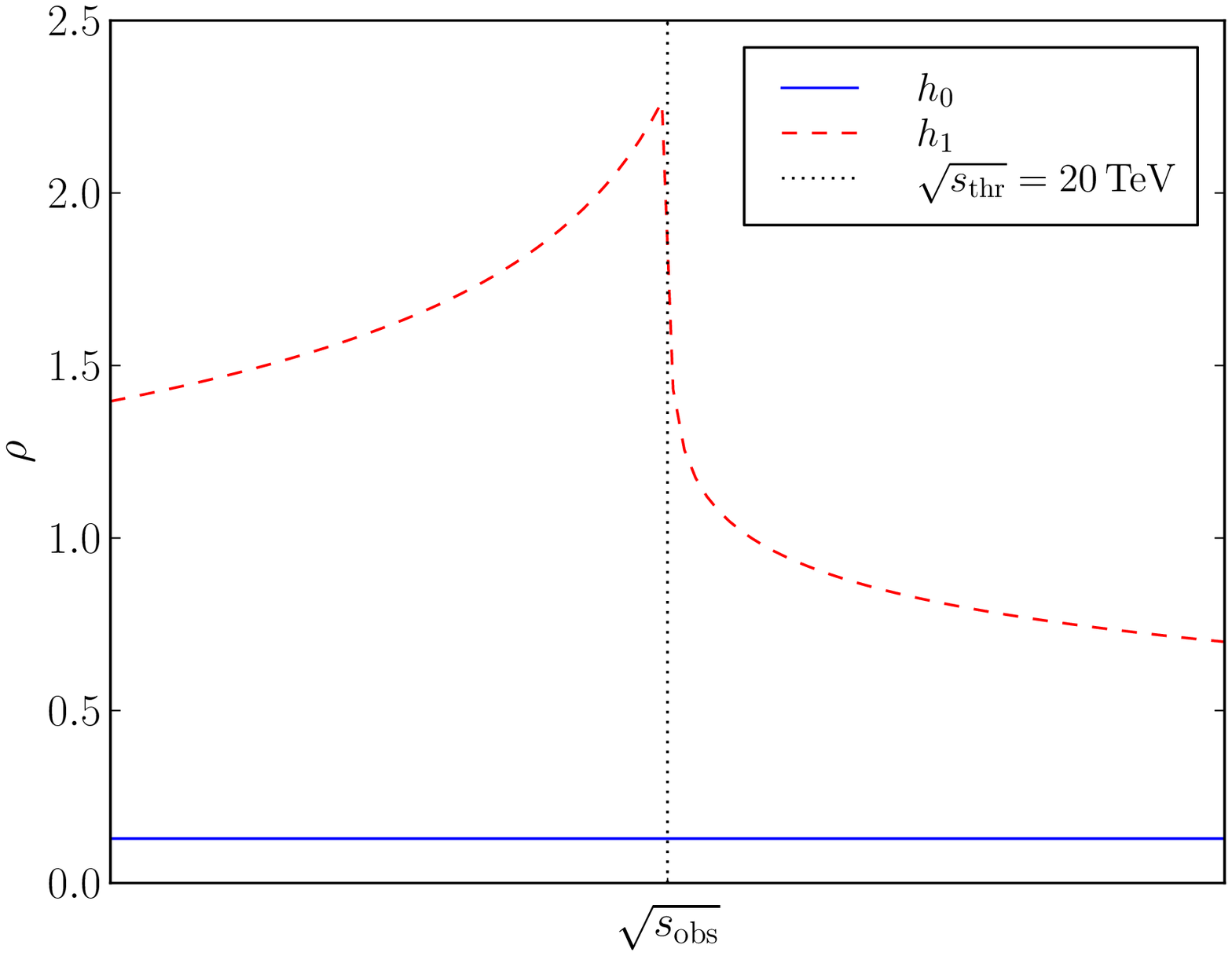}
\caption{At observational energies around LHC energies, the SM $\rho$~(solid line) remains roughly constant. 
Using the step function enhancement $h_1$ with $D=1$ and
$\sqrt{\sthr}=20$ TeV, we find a dramatic increase in $\rho$ well below the new particle threshold. 
The right panel is an expanded piece of the left panel with a width of $\sim500$ GeV on each side of the threshold energy, better
showing the asymmetry of $\rho$ about its singular peak value.}
\label{fig:rhoVssobstr20}
\end{figure*}
\begin{figure*}[tb]
\centering
\includegraphics[width=\columnwidth]{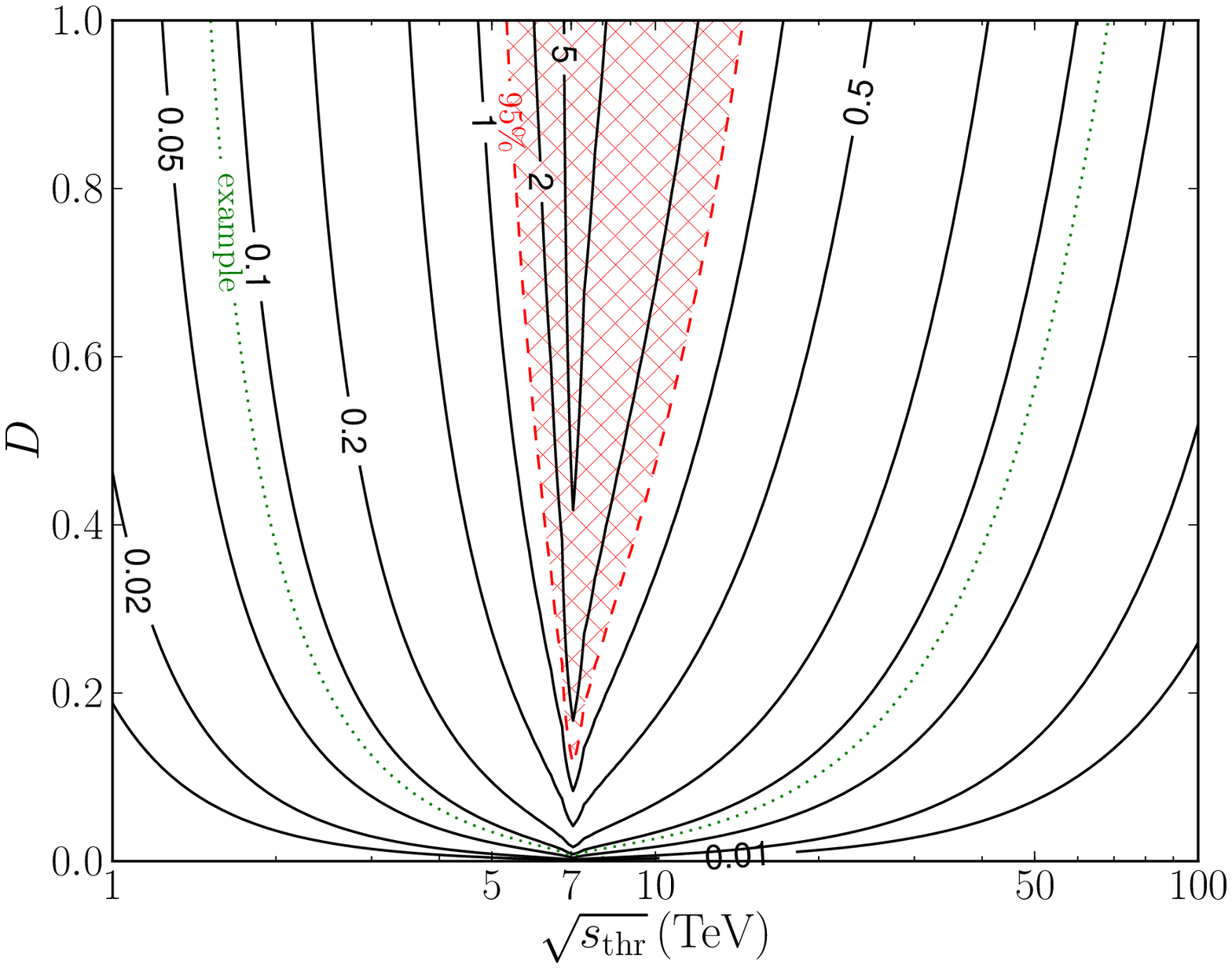}
\includegraphics[width=\columnwidth]{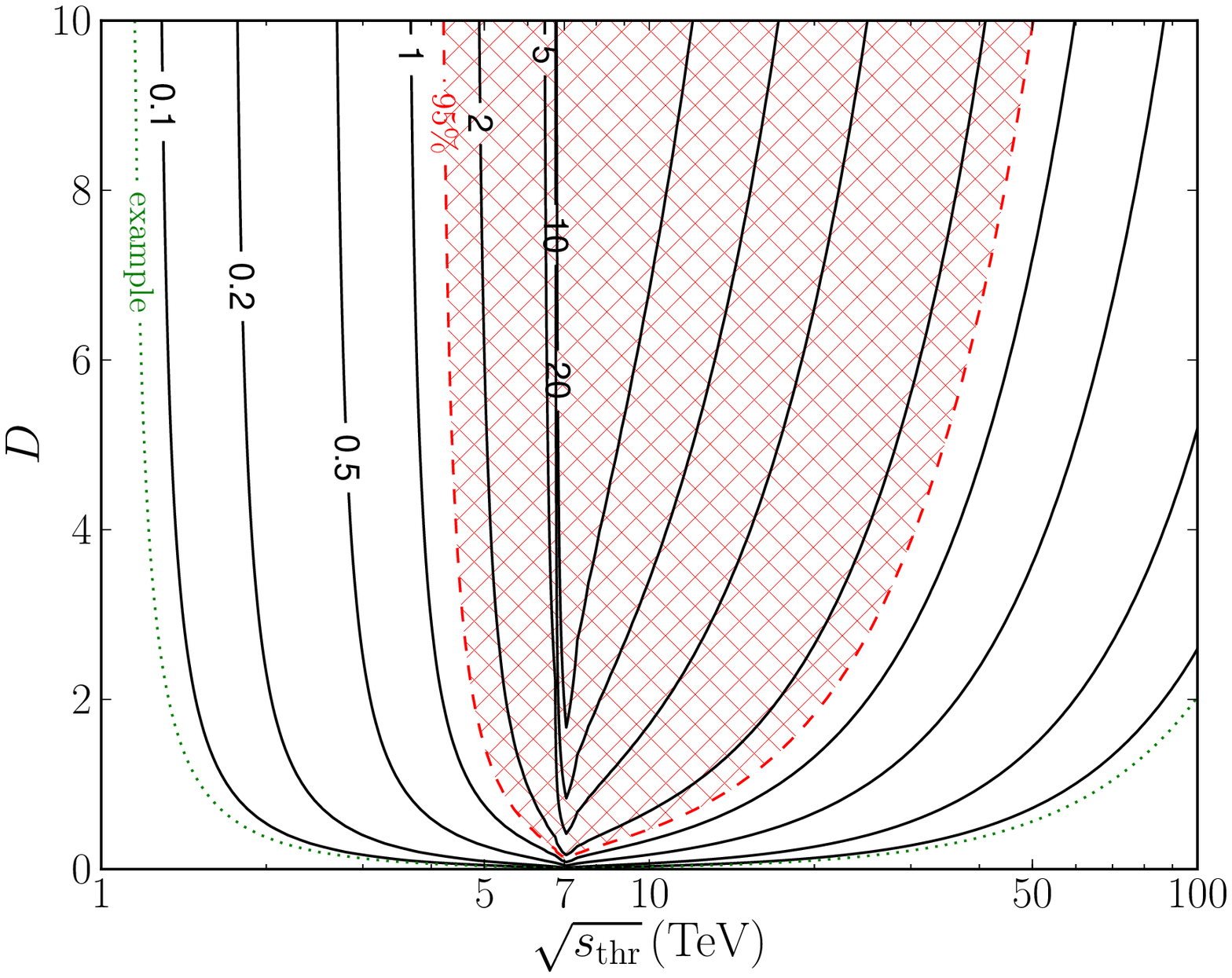}
\caption{At $\sqrt{s_{\text{obs}}}=7$ TeV, the contours are parameterized by $\Delta\rho/\rho$.
The form of the enhancement that includes new physics is $h_1$ -- the step function. 
The left panel considers $D$ in the range [0,1], which is the relevant parameter space for SUSY-type models.
The right panel considers $D$ values up to 10, relevant for extra-dimensional theories that have 
arbitrarily large increases in the cross section. The shaded regions have already been ruled out by
TOTEM's $\sqrt s=7$ TeV preliminary results. The dotted green contours correspond to the $\rho=0.145$ example signal.}
\label{fig:strXdXrhoobs7}
\end{figure*}

Next, we look at what range of $\sqrt{\sthr}$ and $D$ values will give an large increase in $\rho$. 
The left and right panels of Fig.~\rf{fig:strXdXrhoobs7}
show contours of $\Delta\rho/\rho$ in the ranges $D\in[0,1]$ and $D\in[0,10]$. 
The $D\in[0,1]$ range of the left panel may be relevant to broken $R$-parity violating SUSY-like models, 
in which  some or all of the SM particles might be doubled. 
The larger $D$ range is plotted in the right panel, to show the increased reach of IDRs
for still larger cross sections, as might be the case with extra-dimensional models.
For the simple case of a step function with a significant increase in cross section, 
we see that IDRs offer a very powerful window to physics BSM.

Also displayed in Fig.~\rf{fig:strXdXrhoobs7} are the regions of the generous $h_1$ step function model 
that are ruled out at 95\% significance by these TOTEM results.
The IDR technique is sensitive to a large range of $(\sqrt{\sthr},D)$ parameter space of the $h_1$ step function model,
even with the currently large TOTEM errors on the IDR-independent $\rho$.
In particular, the IDR technique is sensitive to new energy thresholds well beyond the direct energy reach of the LHC. 
A minimal inference to be drawn from the 95\% confidence level exclusion in the figure is that  
the cross section cannot increase particularly quickly near the LHC energy $\sqrt s=7$~TeV\@.

Each higher energy probed by the LHC will rule out an additional region of $h_1$ parameter space.
Going forward, improvements are planned for the TOTEM optics, which will reduce the errors on $\rho$ 
and thereby increase the sensitivity of the IDR toolkit to BSM physics.

Shown also in Fig.~\rf{fig:strXdXrhoobs7} is the contour corresponding to our $\rho=0.145$ example signal.
Our example value of $\rho$ is taken from the TOTEM experiment's inferred mean value. 
If such a signal were statistically and systematically significant, 
we would expect new physics to show up as an increase in the $pp$ cross section of height $D$ and 
threshold $\sqrt{\sthr}$ somewhere on this contour. 
(We don't consider a signal of new physics at energies much below the machine energy, 
as direct detection of new event topologies or increased cross section would likely provide a better signal 
than a change in $\rho$ as inferred through IDRs.)

\quad

We now turn to our more realistic models, $h_2$ and $h_3$, describing the onset of new physics.
The $h_1$ model contains two parameters, $D$ and $\Ethr$, and so for this model 
we showed the prediction for $\Delta\rho/\rho$ as a contour plot.
With the $h_2$ and $h_3$ models, there is no analog of $D$, and the only parameter is $\Ethr$.
Thus, we may show $\Delta\rho/\rho$ and $\Delta\sigma/\sigma$ for these models 
as simple ordinates versus the mass $M_\chi$ of the new particle, and we do so.
With these models, our enthusiasm for the IDR approach will be somewhat tempered.

\subsection{Results From the Partonic Model}
Results for the partonic $h_2$~model are displayed in Fig.~\rf{fig:drhodsigmaVM},
for the next-run LHC energy of $\sqrt{s}=14$~TeV\@.
\begin{figure}
\centering
\includegraphics[width=\columnwidth]{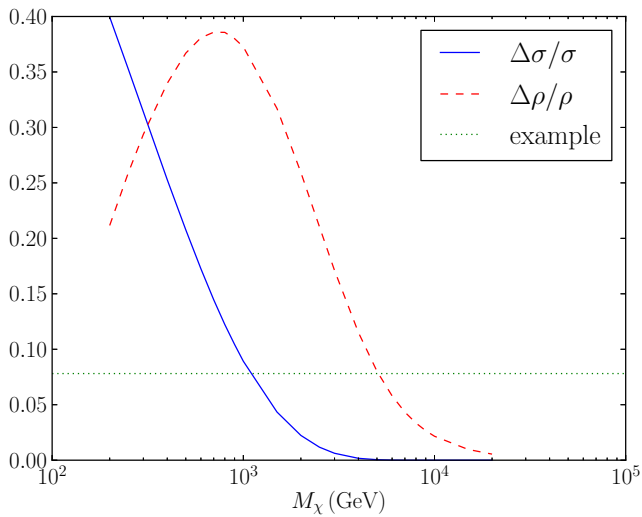}
\caption{The fractional increases in $\rho$ and $\sigma$ using $h_2(s)$ at $\sqrt{s_{\text{obs}}}=7$~TeV, versus $M_\chi$.
With the present significance of $\rho$ data, the exclusion region is well above the top of the graph.
The dotted green line presents the value of $\Delta\rho/\rho$ corresponding to the $\rho=0.145$ example signature; 
from intersecting lines, a new $M_\chi=5.1$~TeV threshold is predicted.}
\label{fig:drhodsigmaVM}
\end{figure}
For a small range of $M_\chi$ values, it is seen that $\Delta\rho/\rho$ is significantly larger than $\Delta\sigma/\sigma$.
However, at present, the errors on the IDR-independent $\rho$ measurement ($\Delta\rho/\rho\alt 1.38$ at $2\sigma$) 
are much larger than the accuracy ($\sim$5\%) 
with which energy-dependent changes in the total cross section can be inferred, so care is warranted here.
From Fig.~\rf{fig:drhodsigmaVM} we can estimate a region of energy in the 2-5~TeV range for which 
$\Delta\rho/\rho\gtrsim0.1$ and $\Delta\sigma/\sigma\lesssim0.05$.
Our inference is that for new particle masses in the $\sim$2-5~TeV energy range, 
IDR-independent measurements of $\rho$ to an accuracy of one part in ten could reveal new physics of the
type described by $h_2$ in section~\ref{subsec:amorephysicalmodel} at $\sqrt{s_{\text{obs}}}=7$~TeV\@. 

One may wonder why the peak in $\rho$ occurs so far below the machine energy of 14~TeV\@.
The reason is the slow rise of the BSM cross section due to suppression from the pdfs:
a peak at energy $\Ethr\sim \sqrt{s}$ weighted by the mean value of the parton momenta product, $\langle x_1 x_2\rangle$,
gives a peak at roughly an order of magnitude below the machine energy.
A second inference is that models with new physics arising from initial state partons 
will enhance the value of $\rho$ mainly below the machine energy.
Of course, such models will also enhance the cross section below the machine energy, as seen in Fig.~\rf{fig:drhodsigmaVM}.

There is still a small increase in $\rho$ at the machine energy of $\sqrt{\sthr}>7$~TeV
due to particle masses beyond 7~TeV\@.
Beyond the machine energy, it is impossible for direct production to occur, so an inference of 
nonzero $\Delta\rho/\rho>0$ due to particle masses beyond $\sqrt{\sthr}>7$~TeV would present a unique, and striking, discovery.
Unfortunately, in the $h_2$~model, such an inference does not seem possible, as $\Delta\rho/\rho$ is $\lesssim0.01$ 
for new particle masses just beyond 7~TeV\@.
A more optimistic inference is that, if the cross section were to rise much more quickly than that of the $h_2$ model,
as happens with a Kaluza-Klein tower of new particles, 
it may be possible to infer such new physics even if the threshold energies/new masses exceed the LHC energy.

We see that our example signal, plotted in Fig.~\rf{fig:drhodsigmaVM}, 
implicates a new mass-scale $M_\chi=5.1$ TeV\@. 
(The example $\Delta\rho/\rho$ also crosses the continuous curve at an energy below the machine energy; 
we assume that any new physics at this lower energy would be detected through more direct means.)
\begin{figure}[tb]
\centering
\includegraphics[width=\columnwidth]{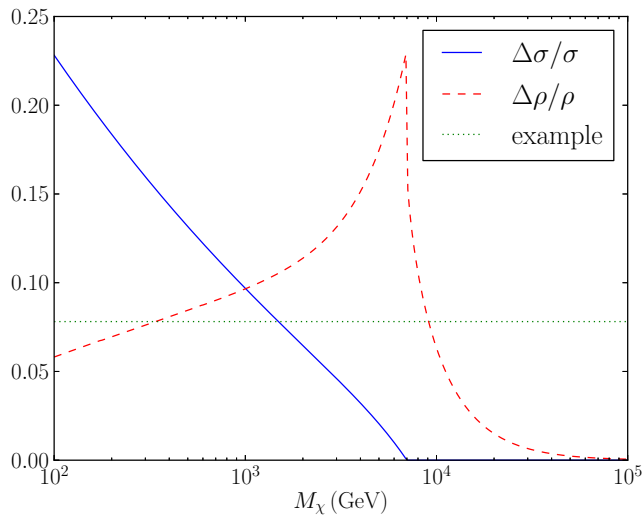}
\caption{The fractional increase in $\rho$ and $\sigma$ for the $h_3$~model, at $\sqrt s=7$~TeV\@.
The increase in $\rho$ compared to its SM value shows a peak of $0.23$ at $M_\chi=\sqrt{\sobs}$.
We also see that $\rho$ has increased compared to its SM value slightly across a range of
energies beyond the observation energy.
With the present significance of $\rho$ data, the exclusion region is well above the top of the graph.
The dotted green line presents the value of $\Delta\rho/\rho$ corresponding to the $\rho=0.145$ example signature; 
from intersecting lines, a new $M_\chi=9.1$~TeV threshold is predicted.}
\label{fig:h3drhodsigma}
\end{figure}

\subsection{Results From the Diffractive Model}
Finally, the $h_3$ model is plotted in Fig.~\rf{fig:h3drhodsigma} at $\sqrt{\sobs}=7$~TeV\@. 
We see a modest contribution to $\rho$ from the $h_3$ modification, as compared to that of the $h_2$~model. 
The larger contribution is due to the faster rise of the $h_3$ model from threshold ($\sthr=M_\chi^2$). 
The non-partonic nature of model $h_3$ is at the heart of the larger, higher-energy peak.
On the other hand, the effect of the smaller increase as $s\to\infty$ as described by Eq.~\rf{eq:h3limit} can be seen
in Fig.~\rf{fig:h3drhodsigma} by the fast fall off in $\Delta\rho/\rho$ beyond $\sqrt{s}=M_{\chi}$.

We note that while no regions of $M_\chi$ parameter space can yet be excluded, our example signal implicates a new 
$M_\chi=9.1$ TeV mass-scale.
(We again ignore the lower energy crossing, where any new physics can be probed in a more direct manner.)
This $\sim 9$~TeV mass-scale has not been directly probed at
the LHC, and likely will be only weakly probed even at the 14~TeV run.

\subsection{Model Conclusions}
In the $h_1$ and $h_3$ models, the peak sensitivity of $\Delta\rho/\rho$ occurs when the new mass/new
physics threshold is right at the machine energy.
The sensitivity then falls off rapidly with increasing mass/threshold. 
However, the phase space for new particle production with mass at the machine energy is zero.
Thus, a cross section measurement will not show an increase for such a mass value.
However, the $\rho$-parameter will show a peak increase.
Thus, the IDR technique primarily extends the reach of the LHC, to particle masses at the very end point of the machine energy.
The LHC discovery potential is also extended beyond the machine energy, but with less sensitivity.
In the $h_2$ model, the parton fractional momenta move the peak sensitivity to lower energies (by about an order of magnitude),
thereby lessening the utility of the IDR technique for extending the LHC discovery potential to the machine energy and beyond.

It appears that this IDR technique may be sensitive to some reasonable models with large changes to the $pp$ cross section, 
which have thresholds exceeding the reach of more direct detection.

The outlook for the near future is dependent on new measurements of $\rho$ from experiments like TOTEM. The $\sqrt s=8$
TeV data from TOTEM is in the process of being analyzed \cite{Kaspar:2013tda}, and we eagerly await the 
next LHC run at $\sqrt s=14$~TeV, which should begin in 2015.

\section*{Acknowledgements}
PBD and TJW are supported in part by Department of Energy grant DE-FG05-85ER40226. 
PBD also acknowledges Vanderbilt University for partial support.

\appendix
\section{IDR Independent Calculation of \texorpdfstring{$\rho$}{rho}}
\label{appx:rhocalculation}
To extract a value for $\rho$ in an IDR-independent fashion,
one invokes the optical theorem and extrapolates $d\sigma/dt$ to $t=0$, as shown below.
The cross section is related to the scattering amplitude by a simple exponential at low $|t|$. The differential cross section is
\[\td{\sigma}t=\frac\pi{k^2}|f|^2\,.\]
At $t=0$ one has
\bea{}
\left.\frac{d\sigma}{dt}\right|_{t=0} &=& \frac\pi{k^2}|\Re f(t=0)+i\Im f(t=0)|^2 \nonumber\\
&=& \frac\pi{k^2}|(\rho+i)\Im f(t=0)|^2\,. \nonumber
\eea
Making use of the optical theorem, \\
$\sigma_{\text{tot}}=(4\pi/k)\Im f(t=0)$, one arrives at 
\[16\pi\left.\td{\sigma}t\right|_{t=0}=(\rho^2+1)\sigma_{\text{tot}}^2\,,\]
where the desired $\rho$ is the ratio of the real and imaginary parts of $f(t=0)$.
From~\cite{Block:1984ru}, the $pp$~differential cross section in the low $t$ limit is well approximated by
\[\td\sigma t\propto e^{Bt}\,,\]
where $B$ is the ``slope parameter", assumed and measured to be very nearly constant.
Thus, a measurement or estimate of $\sigma_{\text{tot}}$ and an extrapolation of $d\sigma/dt$ to $t=0$ via the measured 
slope parameter are sufficient to determine $\rho$ independently  from the IDRs.
While $\sigma_{\text{tot}}$ is often evaluated in the ``luminosity-independent" sense
which includes an estimation of $\rho$, it can also be evaluated (although, less precisely) using a luminosity calculated through
particle counting or beam sweeping techniques.

We also note that since the determination of $\rho$ actually gives a value for $\rho^2$ there is an additional sign ambiguity. There
are two approaches to dealing with this. The first is to compare results from modified cross sections in IDRs to either the positive
or negative values, treating each equally. The second is to note that the IDR results for $\rho$ from all of the fits done to the
$pp$, $p\bar p$ cross sections (regardless of whether or not they follow the Froissart bound) yield a positive value for $\rho$. In
practice we use TOTEM's quoted upper limit on $\rho$ statistically calculated from $\rho^2$ which accounts for the possibility that
$\rho$ could be negative and only places an upper limit on $\rho$.
\section{Minimum Transfer Energy in Light to Light Plus One Heavy Processes}
\label{appx:mint}
We need $|t|$ small in $h_2$ to avoid amplitude suppression by propagators.
Here we calculate the kinematic range of $t$ in the $2\rarr 2$ process
$p_1+p_2\to k_1+k_2$, with $p_1^2=p_2^2=k_2^2=0$ all labeling SM particles and 
$k_1^2=M_\chi^2$ labeling a new heavy particle. 
We will see that $t=0$ is allowed,
leading to an unsuppressed amplitude for massless particle exchange.

Let $\theta$ be the angle between $p_1$ and $k_1$ in the CoM frame.  
Then the  transfer energy squared is
\begin{align*}
t&=(p_1-k_1)^2\\
&=M_\chi^2-2(p^0_1k^0_1-|\vec p_1|\,|\vec k_1|\,\cos\theta)\,,
\end{align*}
where $k^0_1={\vec k_1}^2+M_\chi^2$.
So, $t_{\max/\min}$ are given by
\begin{equation}
t_{\max/\min}= M_\chi^2-2p^0_1k^0_1\pm2|\vec p_1|\,|\vec k_1| \nonumber
\end{equation}
at $\theta=0,\pi$ respectively. Then we have
\[p^0_1=\frac{\sqrt{\hat s}}2\,,\qquad k^0_1=\frac{\hat s+M_\chi^2}{2\sqrt{\hat s}}\,,\quad{\rm and\ }\]
\[|{\vec p_1}|=p^0_1,\qquad |\vec k_1|=\frac{\hat s-M_\chi^2}{2\sqrt{\hat s}}\,,\quad{\rm and\ so\ }\]
\[p^0_1 k^0_1=\frac{\hat s+M_\chi^2}4\,,\qquad |{\vec p_1}|\,|{\vec k_1}|=\frac{\hat s-M_\chi^2}4\,.\]
Then, the maximum/minimum values of $t$ are
\[t=M_\chi^2-\frac{\hat s+M_\chi^2}2\pm\frac{\hat s-M_\chi^2}2=\binom0{M_\chi^2-\hat s}\]
where $t=0$ occurs for the forward scattering $\theta=0$ case, 
and the maximum $|t|$ transfer occurs for the backward scattering $\theta=\pi$ case. 
The $t_{\min}=0$ result confirms that $pp\to\chi+$light\ particles will favor small $|t|$.

\bibliography{bib}

\begin{thebibliography}{10}
\bibitem{note1}%
  \BibitemOpen
  \bibinfo {note} {The $\log^2(s/m_p^2)$ growth is perhaps best revealed by
  first expanding Eq.~(\ref{eq:dsigmadxi}) in $\epsilon$, as
  $\frac{d\sigma}{d\xi}\propto \frac{1+\xi}{\xi}\,(1-\epsilon\,\ln\xi +\cdots)$
  and then integrating.}%
  \bibAnnoteFile{Stop}{note1}%
\end{thebibliography}
\end{document}